\documentclass{article}
\usepackage[draft]{aism-2e,natbib}
\bibpunct{(}{)}{;}{a}{}{,}

\usepackage{graphicx,amsmath}
\usepackage{amsfonts}
\usepackage{amssymb}
\usepackage{graphicx}

\newtheorem{proposition}{\sc Proposition}
\newtheoremrm{definition}{\sc Definition}
\newtheoremrm{assumption}{\sc Assumption}
\newtheoremrm{condition}{\sc Condition}
\newtheoremrm{remark}{\it Remark\/}
\newtheoremrm{example}{\it Example\/}
\newtheoremrm{case}{\it Case\/}
\newtheoremrmno{proof}{\sc Proof}
\newtheoremrmno{procedure}{\sc Procedure}
\newtheoremrmno{note}{\it Note\/}
\newtheoremrmno{fact}{\it Fact\/}
\volumenumber{00}
\vnumber{0}
\frompage{1}
\topage{000}
\publishyear{0000}

\shortauthor{M. R. Grasselli}

\received{xxxx xx, 2000}
\revised{xxxx xx, 2000}
\title{Dual Connections in Nonparametric Classical Information Geometry}

\shorttitle{Nonparametric Information Geometry}

\author{M. R. Grasselli}

\affiliation{Department of Mathematics and Statistics \\
McMaster University \\
1280 Main Street West \\
Hamilton ON L8S 4K1, Canada \\
e-mail: grasselli@math.mcmaster.ca \\}
\abstract{We construct an infinite--dimensional information manifold based on exponential Orlicz spaces without using the notion of  exponential convergence. We then show that convex mixtures of probability densities lie on the same connected component of this manifold, and characterize the class of densities for which this mixture can be extended to an open segment containing the extreme points. For this class, we define an infinite--dimensional analogue of the mixture parallel transport and prove that it is dual to the exponential parallel transport with respect to the Fisher information. We also
define $\alpha$--derivatives and prove that they are convex mixtures
of the extremal $(\pm 1)$-derivatives.}

\begin{document}
\maketitle

\section{Introduction}

Information geometry is the branch of probability theory dedicated to provide families
of probability distributions with differential geometrical structures. One then uses the tools
of differential geometry in order to have a clear and intuitive picture, as well as rigor, in a variety of
practical applications ranging from neural networks to statistical estimation, from mathematical finance to
nonequilibrium statistical mechanics (see \citet{Sollich01}).

It was just over half a century ago that the Fisher information
\begin{equation}
g_{ij}=\int \frac{\partial\log p(x,\theta)}{\partial\theta^i}
\frac{\partial\log p(x,\theta)}{\partial\theta^j}p(x,\theta)dx
\end{equation}
was independently suggested by \citet{Rao45} and \citet{Jeffreys46} as a Riemannian 
metric for a parametric statistical model 
$\left\{p(x,\theta), \theta = (\theta^1,\ldots,\theta^n)\right\}$. The Riemannian geometry of statistical models
was then studied as a mathematical curiosity for some years, with an 
emphasis in the geodesic distances associated with the Levi-Civita connection
for this metric. A greater amount of attention was devoted to
the subject after \citet{Efron75} introduced the concept of statistical curvature, 
pointing out its importance to statistical inference, as well as
implicitly using a new affine connection, which would be known
as the exponential connection. This exponential connection, 
together with another connection, later to be called the mixture connection,
were further investigated by \citet{Dawid75}. 
The work of several years on the geometric aspects of 
parametric statistical models culminated with the masterful account in \citet{Amari85}, 
where the whole finite dimensional differential-geometric
machinery is employed, including a
one-parameter family of $\alpha$-connections, the essential concept of
duality and the notions of statistical divergence, projections and 
minimization procedures. Among the successes of the research at these early
stages one could single out the rigidity of the
geometric structures, such as  the result concerning the uniqueness
of the Fisher metric with respect to monotonicity in \citet{Chentsov82}  and
Amari's result concerning the uniqueness of the $\alpha$-connections
 introduced by invariant statistical divergences. The ideas were then
extensively used in statistics, in particular higher order
asymptotic inference and curved exponential models (see \citet{KassVos97}).

A different line of investigation in Information Geometry 
took off in the nineties: the search for a fully-fledge infinite dimensional manifold of probability
measures. As for motivations for this quest, one had, on the practical side, the need to deal 
with nonparametric models in statistics, where the shape of the
underlying distribution is not assumed to be known. On a more fundamental
level, there was the desire of having parametric statistical
manifolds defined simply as finite dimensional submanifolds of a well defined
manifold of all probability measures on a sample space. The motivating
idea was already in \citet{Dawid75} and was also addressed by
\citet{Amari85}. The first sound mathematical construction, however, is
due to \citet{PistoneSempi95}. Given a probability 
space $(\Omega,{\cal F},\mu)$, they showed how to construct
a Banach manifold ${\cal M}$ of all probability measures 
equivalent to $\mu$. The Banach space used as generalised coordinates was
the Orlicz space $L^{\Phi_1}$, where $\Phi_1$ is an exponential Young 
function.  In a subsequent work,
\citet{PistoneRogantin99} analyzed further properties of this manifold,
in particular the concepts of orthogonality and submanifolds. In Section 3, we review their construction and present an alternative  proof of the main result in \citet{PistoneSempi95}, namely that the collection of covering neighborhoods ${\cal U}_p$ and
charts $e^{-1}_p$ form an affine $C^{\infty}$--atlas for ${\cal M}$. The crux is Proposition \ref{open}, where we show that the image of overlapping neighborhoods under any chart $e^{-1}_p$ is open in the topology of the target space $L^{\Phi_1}$. 

The next step
in this development was the \citet{GibiliscoPistone98} definition of the exponential connection as the
natural connection induced by the use of $L^{\Phi_1}$. These authors then propose a mixture connection
acting on the pretangent bundle $^*T{\cal M}$ and prove that it is dual 
to the exponential connection, in the sense of duality for Banach spaces. They
further define the $\alpha$-connections through generalised 
$\alpha$-embeddings and show that the formal relation between the exponential,
mixture and $\alpha$-connections are the same as in the parametric case, that
is
\begin{equation}
\nabla^{(\alpha)}=
\frac{1+\alpha}{2}\nabla^{(e)}+\frac{1-\alpha}{2}\nabla^{(m)}.
\label{convex1}
\end{equation} 

We argue, however, that neither of these two results 
(duality for the exponential and mixture connection and $\alpha$-connections
as convex mixture of them) is a proper generalization of the corresponding
parametric ones, the reason being twofold. First, Banach space duality
is not Amari-Nagaoka duality. The latter refers to a metric being preseverd by
the joint action of two parallel transports, which are then said to be dual
(see (\ref{dual})). Secondly, all the $\alpha$-connections in the parametric 
case act on the tangent bundle, whereas in \citet{GibiliscoPistone98} each
of them acts on its own bundle-connection pair, 
making a formula like (\ref{convex1}) at least difficult to interpret.

In order to address these problems, we define in Section 4 an isomorphism $\tau^{(-1)}$  of tangent spaces, which satisfy the Amari--Nagaoka duality relation with respect to the Fisher metric when paired with the exponential parallel transport $\tau^{(1)}$. However, it turns out that our map $\tau^{(-1)}$ can only be rigorously defined between points $q_1$ and $q_2$ in ${\cal M}$ whose ratio is a bounded random variable. Proposition \ref{extended_mix} then characterizes the extended convex mixtures between such points. 

In Section 5, we rearrange the definitions of \citet{GibiliscoPistone98} in order to
have $\alpha$--derivatives all acting on the same tangent bundle, but defined only for a restricted class of tangent vectors. We then show that the desired relation \eqref{convex1} holds for our definitions. We then finalize the paper by showing that the $\alpha$--auto--parallel curves between two points whose ratio is a bounded function belong to the connected component ${\cal E}(p)$. 

\section{Orlicz Spaces}

\addtocounter{equation}{2}

We present here the aspects of the theory of Orlicz spaces that will be 
relevant for the construction of the information manifold. For more comprehensive accounts, as well as for the proofs of all statements in this section, the reader is 
referred to the monographs of \citet{RaoRen91} and \citet{KrasnoselskiiRutickii61}.

The general theory of Orlicz spaces is developed around the concept of a {\em Young} function, that is, a convex function 
$\Phi:\mathbb{R} \mapsto \overline{\mathbb{R}}^{\scriptscriptstyle{+}}$
satisfying 
\begin{enumerate}
\item $\Phi(x)=\Phi(-x), \quad x\in \mathbb{R}$,
\item $\Phi(0)=0$,
\item $\displaystyle{\lim_{x \mapsto \infty}} \Phi (x)= +\infty$. 
\end{enumerate}
For applications in information geometry, it is enough to consider Young functions of the form
\begin{equation}
\label{Nfunctions}
\Phi(x)=\int_0^{|x|} \phi(t)dt, \quad x\geq 0,
\end{equation}
where $\phi:[0,\infty)\mapsto [0,\infty)$ 
is nondecreasing, continuous and such that $\phi(0)=0$ and
$\displaystyle{\lim_{x\rightarrow\infty}}\phi(x)=+\infty$. Young functions of this type include the monomials $|x|^r/r$, for $1<r<\infty$, and the following examples arising in information geometry:
\begin{eqnarray}
\Phi_1(x) &=& \cosh x -1, \\
\Phi_2(x) &=& e^{|x|}-|x|-1, \\
\Phi_3(x) &=& (1+|x|)\log(1+|x|)-|x|
\end{eqnarray}
(in the sequel, $\Phi_1$,$\Phi_2$ and $\Phi_3$ will always refer to these 
three particular functions, with other symbols being used to denote generic 
Young functions).

When a Young function $\Phi$ is given in the form \eqref{Nfunctions} we can define its complementary (conjugate) function  as the Young function $\Psi$ given by
\begin{equation}
\Psi(y)=\int_0^{|y|} \psi(t)dt, \quad y\geq 0,
\end{equation}
where $\psi$ is the inverse of $\phi$. One can verify that $(\Phi_2, \Phi_3)$ and $(|x|^r/r,|x|^s/s)$, with $r^{-1}+s^{-1}=1$,  are examples of complementary pairs. For a general Young function $\Phi$, the complementary function $\Psi$ is given less constructively by 
\begin{equation}
\Psi(y)=\sup\{x\geq 0 : x|y|-\Phi(x)\}.
\end{equation}

There are many different ways of introducing a partial order on the class of Young functions. A particularly straightforward one is to say that a Young function $\Psi_2$ is  {\em stronger} than
another Young function $\Psi_1$, denoted by $\Psi_1\prec\Psi_2$, if there exist a constant $a>0$ such that 
\begin{equation}
\Psi_1(x)\leq\Psi_2(ax), \quad x\geq x_0,
\end{equation}
for some $x_0\geq 0$ (depending on $a$). For example, one can verify that 
\begin{equation}
\label{order}
|x|\prec\Phi_3\prec\frac{|x|^r}{r}\prec\frac{|x|^s}{s}\prec\Phi_2
\end{equation}
whenever $1<r\leq s<\infty$. Two Young functions $\Psi_1$ and $\Psi_2$ are said to be {\em equivalent} if 
$\Psi_1\prec\Psi_2$ and $\Psi_2\prec\Psi_1$, that is, if there exist real numbers $0<c_1 \leq c_2 < \infty$ and $x_0\geq 0$ such that
\begin{equation}
\Psi_1(c_1x)\leq\Psi_2(x)\leq\Psi_1(c_2x), \quad x\geq x_0.
\end{equation}
For example, the functions $\Phi_1$ and $\Phi_2$ are equivalent, both being of exponential type. 

Now let $(\Omega,\Sigma,P)$ be a probability space. The {\em Orlicz class} associated with a Young function $\Phi$
is defined as
\begin{equation}
\tilde{L}^{\Phi}(P)=\left\{f:\Omega\mapsto\overline{\mathbb{R}}, 
\mbox{measurable : }\int_{\Omega}\Phi(f) dP< \infty\right\}.
\end{equation}
Since $P$ is a finite measure, the Banach space $L^\infty(\Omega,\Sigma,P)$ of essentially bounded random variables is easily seen to be a subset of $\tilde{L}^\Phi(P)$ for any Young function $\Phi$. It is easy to see that $\tilde{L}^{\Phi}(P)$ is a convex set and that 
$h\in\tilde L^{\Phi}(P)$ and $|f|\leq |h|$ imply that $f\in \tilde L^{\Phi}(P)$. However, in general, $\tilde{L}^{\Phi}(P)$ is {\em not} a vector space, which leads to the definition of the {\em Orlicz space} associated with a Young function $\Phi$ as
\begin{equation}
L^{\Phi}(P)=\left\{f:\Omega\mapsto\overline{\mathbb{R}}, 
\mbox{measurable : } 
\int_{\Omega}\Phi(\alpha f)dP < \infty, \mbox{ for some } \alpha>0\right\},
\end{equation}
furnished with the {\em Luxembourg} norm (see \citet[page 67]{RaoRen91})
\begin{equation}
N_{\Phi}(f)=\inf\left\{k>0:\int_{\Omega}\Phi\left(\frac{f}{k}\right)dP
\leq 1\right\}.
\end{equation}
or with the equivalent {\em Orlicz} norm (see \citet[page 61]{RaoRen91})
\begin{equation}
\|f\|_{\Phi}=\sup\left\{\int_{\Omega} |fg|dP : g \in L^{\Psi}(P),
 \int_{\Omega}\Psi(g)dP
\leq 1\right\},
\end{equation}
where $\Psi$ is the complementary Young function to $\Phi$. We observe for later use that $\int_{\Omega}\Phi(f)dP 
\leq 1$ iff $N_{\Phi}(f) \leq 1$ (see \citet[page 54]{RaoRen91}).

A key ingredient in the analysis of Orlicz spaces is the generalized 
H\"{o}lder
inequality (see \citet[page 58]{RaoRen91}). If $\Phi$ and $\Psi$ are complementary Young functions, 
$f \in L^{\Phi}(P)$, $g \in L^{\Psi}(P)$, then
\begin{equation}
\int_{\Omega} |fg|dP \leq 2N_{\Phi}(f)N_{\Psi}(g).
\end{equation}
It follows that each element $f\in L^{\Phi}(P)$ defines a continuous linear functional on $L^{\Psi}(P)$, so that if we denote its topological dual by $\left(L^{\Psi}\right)^{*}$ we obtain the continuous injection $L^{\Phi} \subset \left(L^{\Psi}\right)^{*}$ for any 
pair of complementary Young functions. 

If $\Psi_2\prec\Psi_1$ then there exist a constant $k$ such that $N_{\Psi_2}(\cdot)\leq k N_{\Psi_1}(\cdot)$ and 
therefore $L^{\Psi_1}(P)\subset L^{\Psi_2}(P)$ (see \citet[page 155]{RaoRen91}). For instance, due to \eqref{order} we obtain that for $1<r\leq s<\infty$
\begin{equation}
L^{\Phi_2}\subset L^s \subset L^r \subset L^{\Phi_3} \subset L^1,
\end{equation}
where $L^r, r\geq 1$ denote the usual Lebesgue spaces on $(\Omega,\Sigma,P)$, which coincide with the Orlicz space 
defined by the Young functions $|x|^r/r, r\geq 1$. 
If two Young functions are equivalent, then the Orlicz spaces associated with them are isomorphic, that is, they coincide as sets and have equivalent norms. For example, we have that $L^{\Phi_1}(P)= L^{\Phi_2}(P)$.

\section{The Pistone-Sempi Information Manifold}

\addtocounter{equation}{17}

We start by reviewing the construction of an infinite dimensional information manifold along the lines of \citet{PistoneSempi95,
PistoneRogantin99,GibiliscoPistone98}. Consider the set ${\cal M}$ of all densities of probability measures 
equivalent to a reference measure $\mu$, that is,
\[{\cal M}\equiv{\cal M}(\Omega,\Sigma,\mu)=\{f:\Omega\mapsto\mathbb{R},
\mbox{measurable : }
f>0 \mbox{ a.e. and  } \int_{\Omega}fd\mu=1 \}.\]

For each point $p\in{\cal M}$, let $L^{\Phi_1}(p)$ be the exponential 
Orlicz space with norm $N^{\Phi_1}_p(\cdot)$ over the probability space $(\Omega,\Sigma,pd\mu)$ 
and consider its closed subspace of $p$-centred random variables
\begin{equation}
B_p=\{u\in L^{\Phi_1}(p):\int_{\Omega}upd\mu=0\}
\end{equation}
as the coordinate Banach space. 

In probabilistic terms, the set $L^{\Phi_1}(p)$ corresponds to random variables whose
moment generating function with respect to the probability $p d\mu$ is finite on a neighborhood of the origin (see \citet[proposition 2.3]{PistoneSempi95}). In statistics this are exactly the random variables used to define the one dimensional exponential model
$p(t)$ associated with a point $p\in{\cal M}$ and a random variable $u$:
\begin{equation}
p(t)=\frac{e^ {t u}}{Z_p(tu)}p, \qquad t\in(-\varepsilon,\varepsilon).
\end{equation}
In particular, if we denote by ${\cal V}_p$ the unit ball in $B_p$, then it follows that the moment generating functional 
$Z_p(u)=\int_{\Omega}e^upd\mu$
is finite on ${\cal V}_p$ (see \citet[proposition 2.4]{PistoneSempi95}). The underlying idea for the Pistone--Sempi manifold is to
parametrize the neighborhoods around points $p\in{\cal M}$ by all possible one dimensional exponential models passing through $p$. As a preliminary result, we mention that if two densities $p$ and $q$ are connected by a one dimensional exponential model, then 
$L^{\Phi_1}(p)=L^{\Phi_1}(q)$ (see \citet[proposition 5]{PistoneRogantin99}).

Pistone and Sempi define the inverse of a local chart around $p\in {\cal M}$ as 
\begin{eqnarray}
e_p:{\cal V}_p &\rightarrow& {\cal M} \nonumber \\
u &\mapsto& \frac{e^u}{Z_p(u)}p.
\end{eqnarray}

Denote by ${\cal U}_p$ the image of ${\cal V}_p$ under $e_p$. We verify that
$e_p$ is a bijection from ${\cal V}_p$ to ${\cal U}_p$, since
\[\frac{e^u}{Z_p(u)}p=\frac{e^v}{Z_p(v)}p\]
implies that $(u-v)$ is a constant random variable, which must vanish, since both $u,v$ have zero $p$--expectation.
Then let $e^{-1}_p$ be the inverse of $e_p$ 
on ${\cal U}_p$.
One can check that
\begin{eqnarray}
e^{-1}_p:{\cal U}_p &\rightarrow& B_p \nonumber \\
q &\mapsto& \log\left(\frac{q}{p}\right)-\int_{\Omega}\log\left(
\frac{q}{p}\right)pd\mu.
\end{eqnarray}
and also that, for any $p_1,p_2 \in {\cal M}$, the transition functions are given by
\begin{eqnarray}
e^{-1}_{p_2}e_{p_1}:e^{-1}_{p_1}({\cal U}_{p_1}\cap{\cal U}_{p_2})
 &\rightarrow& e^{-1}_{p_2}({\cal U}_{p_1}\cap{\cal U}_{p_2}) \nonumber \\
u &\mapsto& u+\log\left(\frac{p_1}{p_2}\right)-\int_{\Omega}\left(u+\log
\frac{p_1}{p_2}\right)p_2d\mu. \label{transition}
\end{eqnarray}

The main result of \citet{PistoneSempi95} is to show that the charts defined above lead to a well--defined infinite dimensional manifold. The crucial part of the proof is to show that, for any two points $p_1,p_2\in {\cal M}$, the image of the overlapping neighborhoods ${\cal U}_{p_1}\cap{\cal U}_{p_2}$ under $e^{-1}_{p_1}$ is open in the topology of the model space $B_{p_1}$. To do so they introduce a topology induced by the notion of exponential convergence, with respect to which the sets ${\cal U}_{p_1}\cap{\cal U}_{p_2}$ are open, and then show that $e^{-1}_{p_1}$ is sequentially continuous from exponential convergence to $L^{\Phi_1}$--convergence. In what follows, we bypass the use of exponential convergence and present a direct proof that the Pistone and Sempi construction yields a Banach manifold. We first need to establish the following proposition.

\begin{proposition} For any $p_1,p_2 \in {\cal M}$, the set $e^{-1}_{p_1}({\cal U}_{p_1}\cap{\cal U}_{p_2})$ is open
in the topology of $B_{p_1}$.
\label{open}
\end{proposition}
{\it Proof}: Suppose that $q \in {\cal U}_{p_1}\cap{\cal U}_{p_2}$ for 
some $p_1,p_2 \in {\cal M}$. Then we can write it as
\[q=\frac{e^u}{Z_{p_1}(u)}p_1,\]
for some $u\in{\cal V}_{p_1}$. Using \eqref{transition}, we find
\[e^{-1}_{p_2}(q)=u+\log\left(\frac{p_1}{p_2}\right)-\int_{\Omega}\left(u+\log
\frac{p_1}{p_2}\right)p_2d\mu.\]
Since $e^{-1}_{p_2}(q)\in {\cal V}_{p_2}$, we have that
\[N^{\Phi_1}_{p_2}\left(e^{-1}_{p_2}(q)\right)=N^{\Phi_1}_{p_2}\left(u+\log\left(\frac{p_1}{p_2}\right)
-\int_{\Omega}\left(u+\log\frac{p_1}{p_2}\right)p_2d\mu\right)<1.\]
Consider an open  ball of radius $r$ around $u=e^{-1}_{p_1}(q)\in e^{-1}_{p_1}
({\cal U}_{p_1}\cap{\cal U}_{p_2})$ in the topology of $B_{p_1}$, that
is, consider the set
\[A_r=\{v\in B_{p_1} : N^{\Phi_1}_{p_1}(v-u)<r\}\]
and let $r$ be small enough so that $A_r \subset {\cal V}_{p_1}$. Then the 
image in ${\cal M}$ of each point 
$v \in A_r$ under $e_{p_1}$ is
\[\tilde{q}=e_{p_1}(v)=\frac{e^v}{Z_{p_1}(v)}p_1.\]
We claim that $\tilde{q} \in {\cal U}_{p_1}\cap{\cal U}_{p_2}$ if $r$ is sufficiently small. 
Indeed, applying
$e^{-1}_{p_2}$ to it we find 
\[e^{-1}_{p_2}(\tilde{q})=v+\log\left(\frac{p_1}{p_2}\right)-\int_{\Omega}\left(v+\log
\frac{p_1}{p_2}\right)p_2d\mu,\]
so
\begin{eqnarray*}
N^{\Phi_1}_{p_2}\left(e^{-1}_{p_2}(\tilde{q})\right) &\leq& N^{\Phi_1}_{p_2}(v-u)+
N^{\Phi_1}_{p_2}\left(u+\log\left(\frac{p_1}{p_2}\right)
-\int_{\Omega}\left(u+\log\frac{p_1}{p_2}\right)p_2d\mu\right)\\
& & + N^{\Phi_1}_{p_2}\left(\int_{\Omega}(v-u)p_2d\mu\right)\\
&\leq& N^{\Phi_1}_{p_2}(v-u)+N^{\Phi_1}_{p_2}\left(e^{-1}_{p_2}(q)\right)+ 
N^{\Phi_1}_{p_2}(1)\int_{\Omega}|v-u|p_2d\mu\\
&=& N^{\Phi_1}_{p_2}(v-u)+N^{\Phi_1}_{p_2}\left(e^{-1}_{p_2}(q)\right)+\|v-u\|_{1,p_2}K,
\end{eqnarray*}
where $K=N^{\Phi_1}_{p_2}(1)$ and we use the notation $\|\cdot\|_{1,p_2}$ for
the $L^1(p_2)$-norm. As we have seen in the previous section, it follows from the growth properties of $\Phi_1$ that
there exists $c_1>0$ such that $\|f\|_{1,p_2} \leq c_1N^{\Phi_1}_{p_2}(f)$. 
Moreover, since $L^{\Phi_1}(p_1)=L^{\Phi_1}(p_2)$ (since both
$p_1$ and $p_2$ are connected to $q$ by one dimensional exponential models) it follows that there exists a constant 
$c_2>0$ such that  $N^{\Phi_1}_{p_2}(f) \leq c_2N^{\Phi_1}_{p_1}(f)$. Therefore, the 
previous inequality becomes
\begin{eqnarray*}
N^{\Phi_1}_{p_2}\left(e^{-1}_{p_2}(\tilde{q})\right) &\leq& c_2N^{\Phi_1}_{p_1}(v-u)+
N^{\Phi_1}_{p_2}\left(e^{-1}_{p_2}(q)\right)+c_1c_2KN^{\Phi_1}_{p_1}(v-u) \\
&=& c_2(1+c_1K)N^{\Phi_1}_{p_1}(v-u)+N^{\Phi_1}_{p_2}\left(e^{-1}_{p_2}(q)\right).
\end{eqnarray*}
Thus, if we choose 
\[r< \frac{1-N^{\Phi_1}_{p_2}\left(e^{-1}_{p_2}(q)\right)}{c_2(1+c_1K)},\]
we will have that 
\[N^{\Phi_1}_{p_2}\left(e^{-1}_{p_2}(\tilde q)\right)<1\]
which proves the claim. What we have just proved is that 
\mbox{$e^{-1}_{p_1}({\cal U}_{p_1}\cap{\cal U}_{p_2})$} 
consists entirely of interior points in the 
topology of $B_{p_1}$, and is therefore 
open in $B_{p_1}$.
\vspace{0.2in}

We then have that the collection $\{({\cal U}_p,e^{-1}_p),p\in {\cal M}\}$ 
satisfies
the three axioms for being a $C^{\infty}$--atlas for ${\cal M}$ 
(see \citet[p 20]{Lang95}). Moreover, since for each connect component all the spaces $B_p$ are 
isomorphic as topological vector spaces, we can say that ${\cal M}$ is a $C^{\infty}$--manifold 
modeled on $B_p$.

As usual, the tangent space at each point $p\in {\cal M}$ can be abstractly 
identified with $B_p$. A concrete realisation has been given in 
\citet[proposition 21]{PistoneRogantin99}, namely each curve 
through $p\in {\cal M}$ is tangent to a one-dimensional exponential model
$\frac{e^{tu}}{Z_p(tu)}p$, so we take $u$ as the tangent vector representing
 the 
equivalence class of such a curve.  

Finally, given a point $p\in {\cal M}$, the connected component of ${\cal M}$ containing $p$ coincides with the
{\em 
maximal exponential model} obtained from $p$ (see \citet[theorem 4.1]{PistoneSempi95}):
\begin{equation}
{\cal E}(p)=\left\{\frac{e^u}{Z_p(u)}p, u \in B_p\cap {\cal Z}_p \right\},
\end{equation}
where ${\cal Z}_p=\{f:Z_p(f)<\infty\}^0$.

\section{The Fisher Information and Dual Connections}

\addtocounter{equation}{23}

In the parametric version of information geometry, Amari and Nagaoka have 
introduced the concept of dual connections with respect to a Riemannian
metric (see \citet{AmariNagaoka00} and the references given therein to their earlier work). 
For finite dimensional manifolds, any continuous assignment of a
positive definite symmetric bilinear form to each tangent space determines
a Riemannian metric. In infinite dimensions, we need to impose that the 
tangent space be self-dual and that the bilinear form be bounded. Since
our tangent spaces $B_p$ are not even reflexive, let alone self-dual, we 
abandon the idea of having a Riemannian structure on ${\cal M}$ and 
propose a weaker version of duality, the duality with respect to a continuous 
scalar product. When restricted to finite dimensional submanifolds, the scalar
product becomes a Riemannian metric and the original definition of duality is
recovered.

Let $\langle \cdot,\cdot \rangle_p$ be a continuous positive definite symmetric
bilinear form assigned continuously to each $B_p\simeq T_p{\cal M}$. 
A pair of connections $(\nabla,\nabla^*)$ are said to be dual with respect to 
$\langle \cdot,\cdot \rangle_p$ if
\begin{equation}
\label{dual_def}
\langle \tau u,\tau^*v \rangle_q=\langle u,v \rangle_p
\end{equation}
for all $u,v \in T_p{\cal M}$ and all smooth curves $\gamma:[0,1]\rightarrow
{\cal M}$ such that $\gamma(0)=p$,$\gamma(1)=q$, where $\tau$ and $\tau^*$ 
denote 
the parallel transports associated with $\nabla$ and $\nabla^*$, respectively. 
Equivalently, $(\nabla,\nabla^*)$ are  dual with respect to 
$\langle \cdot,\cdot \rangle_p$ if
\begin{equation}
v\left(\langle s_1,s_2 \rangle_p\right)=\langle \nabla_vs_1,s_2 \rangle_p
+ \langle s_1,\nabla^*_vs_2 \rangle_p
\label{dual}
\end{equation}
for all $v\in T_p{\cal M}$ and all smooth vector fields $s_1$ and $s_2$.

We stress that this {\em is not} the kind of duality obtained when a 
connection $\nabla$ on a bundle ${\cal F}$ is used to construct
another connection $\nabla'$ on the dual bundle ${\cal F}^*$ 
as defined, for instance, in \citet[definiton 6]{GibiliscoPistone98}. The 
latter is a construction that does not involve any metric or scalar product 
and the two connections act on different bundles, while Amari-Nagaoka duality
is a duality with respect to a specific scalar product (or metric, in the
finite dimensional case) and the dual connections act on the same bundle,
the tangent bundle.

The infinite dimensional generalisation of the Fisher information is given by
\begin{equation}
\langle u,v \rangle_p=\int_{\Omega}(uv)pd\mu, \quad \forall u,v \in B_p.
\end{equation}
This is clearly bilinear, symmetric and positive definite. Moreover, continuity follows from that fact that, since 
$L^{\Phi_1}(p)= L^{\Phi_2}(p) \subset L^{\Phi_3}(p)$, the generalized H\"{o}lder inequality gives
\begin{equation}
|\langle u,v \rangle_p| \leq K N^{\Phi_1}_p(u) N^{\Phi_1}_p(v), 
\quad \forall u,v \in B_p.
\end{equation}

The use of exponential Orlicz space to model the manifold naturally induces 
a globally flat affine connection on the tangent bundle $T{\cal M}$, 
called the 
{\em exponential} connection and denoted by $\nabla^{(1)}$. It is defined on 
each connected component of the manifold ${\cal M}$, which is equivalent
to saying that its parallel transport is defined between points connected by an
exponential model. If $q_1$ and $q_2$ are 
two such points, then the exponential
parallel transport is given by
\begin{eqnarray}
\tau_{q_1q_2}^{(1)}:  T_{q_1}{\cal M} &\rightarrow&  T_{q_2}{\cal M} \nonumber \\
u &\mapsto& u-\int_{\Omega}uq_2d\mu.
\label{exp_parallel}
\end{eqnarray}
It is a well--defined isomorphism, since $T_{q_1}{\cal M}=B_{q_1}$ and $T_{q_2}{\cal M}=B_{q_2}$ are subsets
of the same Orlicz space $L^{\Phi_1}(q_1)=L^{\Phi_1}(q_2)$, so the 
exponential parallel transport just subtracts a constant from $u$ to make it
centred around the right point. 
 
We now want to obtain the dual parallel transport to $\tau^{(1)}$ with respect 
to the Fisher information, which in the parametric version of information geometry is called the mixture parallel transport since it is derived from  the convex mixture of two densities. We therefore start with a result regarding such mixtures.  

\begin{proposition} If $q_1$ and $q_2$ are two points in ${\cal U}_p$
for some $p \in {\cal M}$, then 
\[q(t)=tq_1 + (1-t)q_2\] belongs to ${\cal E}(p)$ for all $t \in [0,1]$.
\label{mixtureg}
\end{proposition}
{\em Proof:} We begin by writing
\[ q_1=\frac{e^{u_1}}{Z_p(u_1)}p \quad \mbox{and} \quad
 q_2=\frac{e^{u_2}}{Z_p(u_2)}p,\]
for some $u_1,u_2 \in {\cal V}_p \subset L^{\Phi_1}(p)$. Therefore, there exist constants $\beta_1>1$ and $\beta_2>1$ such that
$\int_\Omega \Phi_1(\beta_1 u_1)pd\mu <\infty$ and $\int_\Omega \Phi_1(\beta_2 u_2)pd\mu < \infty$. To simplify the notation, let us define 
\[\tilde{u}_1=u_1-\log Z_p(u_1) \quad \mbox{and} 
\quad \tilde{u}_2=u_2-\log Z_p(u_2).\]
We want 
to show that, if we write 
\[e^{\tilde{u}}p=q(t)=te^{\tilde{u}_1}p+(1-t)e^{\tilde{u}_2}p,\]
then $\tilde{u}$ is an element of ${\cal Z}_p$, so that 
\[u=\tilde{u} - \int_{\Omega}
\tilde{u}pd\mu \in B_p\cap{\cal Z}_p\] and \[q(t)=\frac{e^u}{Z_p(u)}p \in {\cal E}(p).\]

For this, let $\beta=\min(\beta_1,\beta_2)>1$ and observe that, on account of the inequality 
$|a+b|^{\beta}\leq 2^{\beta-1}(|a|^{\beta}+|b|^{\beta})$, we have that 
\begin{eqnarray*}
e^{\beta\tilde{u}} &=& \left|te^{\tilde{u}_1}+(1-t)e^{\tilde{u}_2}\right|^\beta \\
&\leq& 2^{\beta-1}\left(|t|^\beta e^{\beta\tilde{u}_1}+|1-t|^\beta e^{\beta\tilde{u}_2}\right).
\end{eqnarray*}
Thus
\begin{equation}\int_{\Omega}e^{\beta\tilde{u}}pd\mu \leq 2^{\beta-1}|t|^\beta \int_{\Omega}e^{\beta\tilde{u}_1}p
d\mu+2^{\beta-1}|1-t|^\beta\int_{\Omega}e^{\beta\tilde{u}_2}pd\mu < \infty
\label{int1}
\end{equation}
since both $\beta\tilde{u}_1$ and $\beta\tilde{u}_2$ are in $\tilde L^{\Phi_1}(p)$. 
On the other hand, we observe that
\[e^{-\beta\tilde{u}} =\frac{1}{\left(te^{\tilde{u}_1}+(1-t)e^{\tilde{u}_2}\right)^\beta}\leq \frac{1}{t^\beta e^{\beta\tilde{u}_1}}.\]
Therefore
\begin{equation}
\int_{\Omega}e^{-\beta\tilde{u}}pd\mu \leq
 t^{-\beta}\int_{\Omega}e^{-\beta\tilde{u}_1}pd\mu < \infty,
 \label{int2}
  \end{equation}
since $\beta\tilde{u}_1 \in \tilde L^{\Phi_1}(p)$. But this completes the proof, since \eqref{int1} and \eqref{int2} together imply that
$\tilde u \in {\cal Z}_p$.

\vspace{0.2in}

We now explore the possibility of extending the convex mixture between $q_1$ and $q_2$ beyond these extreme points while maintaining positivity of $q(t)$. This depends on the relative sizes of $q_1$ and $q_2$, as shown in the next proposition:

\begin{proposition} Let $q_1=\frac{e^{u_1}}{Z_p(u_1)}p$ and $q_2=\frac{e^{u_2}}{Z_p(u_2)}p$ be two points in ${\cal U}_p$. Then there
exist constants $\varepsilon_1>0$ and $\varepsilon_2>0$ such that $q(t)=[tq_1+(1-t)q_2]\in {\cal E}_p$ for all $t\in(-2\varepsilon_1,1+2\varepsilon_2)$ if and only if 
$(u_1-u_2)\in L^\infty$. Moreover, if $(u_1-u_2)\in L^\infty$, then $L^{\Phi_3}(q_1)=L^{\Phi_3}(q_2)$.
\label{extended_mix}
\end{proposition} 
{\em Proof:} Suppose that $q(t)\in{\cal E}_p$ for all $t\in(-2\varepsilon_1,1+2\varepsilon_2)$. Then since $q(-\varepsilon_1)\geq 0$ we have that 
\[-\varepsilon_1 q_1 + (1+\varepsilon_1)q_2 \geq 0 \quad \Rightarrow \quad \frac{q_1}{q_2}\leq \frac{1+\varepsilon_1}{\varepsilon_1}.\]
Similarly, since $q(1+\varepsilon_2)\geq 0$ we have that 
\[(1+\varepsilon_2)q_1-\varepsilon_2 q_2 \geq 0 \quad \Rightarrow \quad \frac{q_1}{q_2}\geq \frac{\varepsilon_2}{1+\varepsilon_2}.\]
Therefore, the random variable 
\[u_1-u_2=\log\left(\frac{q_1}{q_2}\right)+\int_{\Omega}\log\left(\frac{q_1}{p}\right)pd\mu
-\int_{\Omega}\log\left(\frac{q_2}{p}\right)pd\mu\]
is uniformly bounded from above and below. 

Conversely, if we have that $u_1,u_2\in{\cal V}_p$ with $(u_1-u_2)\in L^\infty$ and  
$K=\|u_1-u_2\|_\infty$, then
\[\frac{Z_p(u_2)}{Z_p(u_1)}e^{-K}\leq\frac{q_1}{q_2}\leq \frac{Z_p(u_2)}{Z_p(u_1)}e^K.\] 
We can then conclude that there exist constants $0<\xi_1<1$ and $\xi_2>1$ such that 
\[\xi_1\leq \frac{q_1}{q_2} \leq \xi_2.\]
Then observe that for $t\leq 0$ we have 
\begin{equation*}
q(t)=\left(t\frac{q_1}{q_2}+(1-t)\right)q_2\geq (t\xi_2+(1-t))q_2.
\end{equation*}
Therefore, provided $\frac{1}{1-\xi_2}<t\leq 0$, the inequality above ensures that $q(t)$ is strictly positive. Using the same notation as in the proof of Proposition \ref{mixtureg}, the same inequality gives 
\begin{eqnarray}
\int_{\Omega}e^{-\beta\tilde u}pd\mu&=&\int_{\Omega}\left(\left(t\frac{q_1}{q_2}+(1-t)\right)\frac{q_2}{p}\right)^{-\beta}pd\mu
\nonumber \\
&\leq&\frac{(Z_p(u_2))^{\beta}}{(t\xi_2+(1-t))^{\beta}}\int_{\Omega}e^{-\beta u_2}pd\mu<\infty.
\label{minusalpha1}
\end{eqnarray}
Similarly, for $t\geq 0$ we have 
\begin{equation*}
q(t)=\left(t\frac{q_1}{q_2}+(1-t)\right)q_2\geq (t\xi_1+(1-t))q_2.
\end{equation*}
Therefore, provided $0\leq t<\frac{1}{1-\xi_1}$, this inequality shows that $q(t)$ is strictly positive and that
\begin{eqnarray}
\int_{\Omega}e^{-\beta\tilde u}pd\mu&=&\int_{\Omega}\left(\left(t\frac{q_1}{q_2}+(1-t)\right)\frac{q_2}{p}\right)^{-\beta}pd\mu
\nonumber \\
&\leq&\frac{(Z_p(u_2))^{\beta}}{(t\xi_1+(1-t))^{\beta}}\int_{\Omega}e^{-\beta u_2}pd\mu<\infty.
\label{minusalpha2}
\end{eqnarray}
Moreover, since the first part of the proof of Proposition \ref{mixtureg} holds provided $q(t)$ is positive, we have that 
$q(t)=tq_1+(1-t)q_2\in {\cal E}_p$ for all $t\in\left(\frac{1}{1-\xi_2},\frac{1}{1-\xi_1}\right)$, which completes the proof for the first statement by setting $\varepsilon_1=\frac{1}{2(\xi_2-1)}$ and $\varepsilon_2=\frac{\xi_1}{2(1-\xi_1)}$.

For the second statement in the Proposition, observe that 
\begin{eqnarray}
\label{q1}
q_1&=& \frac{1}{1+\varepsilon_2}q(1+\varepsilon_2)+\frac{\varepsilon_2}{1+\varepsilon_2}q_2 \\
q_2&=& \frac{1}{1+\varepsilon_1}q(-\varepsilon_1)+\frac{\varepsilon_1}{1+\varepsilon_1}q_1, 
\label{q2}
\end{eqnarray}
for positive densities $q(1+\varepsilon_2)$ and $q(-\varepsilon_1)$. Therefore
\[\int_{\Omega}\Phi_3(\beta v)q_1d\mu < \infty \Longleftrightarrow \int_{\Omega}\Phi_3(\beta v)q_2d\mu < \infty,\]
which implies that $L^{\Phi_3}(q_1)=L^{\Phi_3}(q_2)$. Moreover, equations \eqref{q1} and \eqref{q2} can be used to show that the norms 
$N_{\Phi_3,q_1}(\cdot)$ and $N_{\Phi_3,q_2}(\cdot)$ are equivalent. To see this, let $u\in L^{\Phi_3}(q_1)$ and consider 
$v=\frac{u}{N_{\Phi_3,q_1}(u)}$, so that $N_{\Phi_3,q_1}(v)=1$ and consequently 
\[\int_\Omega \Phi_3(v)q_1d\mu \leq 1.\]
Using \eqref{q1} we see that 
\[\frac{1}{1+\varepsilon_2}\int_\Omega \Phi_3(v)q(1+\varepsilon_2)d\mu+
\frac{\varepsilon_2}{1+\varepsilon_2}\int_\Omega \Phi_3(v)q_2d\mu\leq 1,\]
which implies that 
\begin{equation}
\frac{\varepsilon_2}{1+\varepsilon_2}\int_\Omega \Phi_3(v)q_2d\mu\leq 1.
\label{q3}
\end{equation}
On the other hand, it follows from convexity of $\Phi_3$ that 
\[\Phi_3\left(\frac{\varepsilon_2}{1+\varepsilon_2}v\right)\leq\frac{\varepsilon_2}{1+\varepsilon_2}\Phi_3\left(v\right).\]
Inserting this into \eqref{q3} and denoting $K=\frac{1+\varepsilon_2}{\varepsilon_2}$ gives
\[\int_\Omega\Phi_3\left(\frac{v}{K}\right)q_2d\mu \leq 1,\]
which means that $N_{\Phi_3,q_2}(v)\leq K$. Consequently we have that 
\[N_{\Phi_3,q_2}(u)=N_{\Phi_3,q_2}(v)N_{\Phi_3,q_1}(u)\leq KN_{\Phi_3,q_1}(u).\]
Similarly, let $f\in L^{\Phi_3}(q_2)$ and consider 
$g=\frac{f}{N_{\Phi_3,q_2}(f)}$, so that $N_{\Phi_3,q_2}(g)=1$ and consequently 
\[\int_\Omega \Phi_3(g)q_2d\mu \leq 1.\]
Using \eqref{q2} we see that 
\[\frac{1}{1+\varepsilon_1}\int_\Omega \Phi_3(g)q(-\varepsilon_1)d\mu+
\frac{\varepsilon_1}{1+\varepsilon_1}\int_\Omega \Phi_3(g)q_1d\mu\leq 1,\]
which implies that 
\begin{equation}
\frac{\varepsilon_1}{1+\varepsilon_1}\int_\Omega \Phi_3(g)q_1d\mu\leq 1.
\label{q4}
\end{equation}
Again, it follows from convexity of $\Phi_3$ that 
\[\Phi_3\left(\frac{\varepsilon_1}{1+\varepsilon_1}g\right)\leq\frac{\varepsilon_1}{1+\varepsilon_1}\Phi_3\left(g\right).\]
Inserting this into \eqref{q4} and denoting $k=\frac{1+\varepsilon_1}{\varepsilon_1}$ gives
\[\int_\Omega\Phi_3\left(\frac{g}{k}\right)q_1d\mu \leq 1,\]
which means that $N_{\Phi_3,q_1}(g)\leq k$. Consequently we have that 
\begin{equation}
N_{\Phi_3,q_1}(f)=N_{\Phi_3,q_1}(g)N_{\Phi_3,q_2}(f)\leq kN_{\Phi_3,q_2}(f).
\label{k}
\end{equation}

\vspace{0.2in}

\begin{proposition} Let $q_1=\frac{e^{u_1}}{Z_p(u_1)}p$ and $q_2=\frac{e^{u_2}}{Z_p(u_2)}p$ be two points in ${\cal U}_p$ such that 
$(u_1-u_2)\in L^\infty$. Then the map
\begin{eqnarray}
\tau_{q_1q_2}^{(-1)}:  T_{q_1}{\cal M} &\rightarrow&  T_{q_2}{\cal M} \nonumber \\
u &\mapsto& \frac{q_1}{q_2}u,
\label{mixture_tau}
\end{eqnarray}
is an isomorphism of Banach spaces.  
\label{iso}
\end{proposition}
{\em Proof:} In view of Proposition \ref{extended_mix}, we have that $L^{\Phi_3}(q_1)=L^{\Phi_3}(q_2)$ and that the norms 
$N_{\Phi,q_1}(\cdot)$ and $N_{\Phi,q_2}(\cdot)$ are equivalent, from which it follows that, for
$u\in B_{q_1}$,   
\begin{eqnarray*}
\left\|\frac{q_1}{q_2}u\right\|_{\Phi_1,q_2} &=& \sup\left\{\int_\Omega \left|\frac{q_1}{q_2}uv\right|q_2d\mu : v \in L^{\Phi_3}(q_2), 
\int_{\Omega}\Phi_3(v)q_2d\mu 
\leq 1\right\} \\
&=& \sup\left\{\int_\Omega \left|\frac{q_1}{q_2}uv \right|q_2d\mu : v \in L^{\Phi_3}(q_2),  N_{\Phi_3,q_2}(v)\leq 1\right\} \\
&=& k\sup\left\{\int_\Omega \left|\frac{q_1}{q_2}u\left(\frac{v}{k}\right)\right|q_2d\mu : v \in L^{\Phi_3}(q_2),  N_{\Phi_3,q_2}(v)\leq 1\right\} \\
&\leq& k\sup\left\{\int_\Omega |uf|q_1d\mu : f \in L^{\Phi_3}(q_1), N_{\Phi_3,q_1}(f)\leq 1\right\} \\
&=& k\|u\|_{\Phi_1,q_1}< \infty,
\end{eqnarray*}
where $k$ is the constant appearing in \eqref{k}. Thus, $\frac{q_1}{q_2}u \in L^{\Phi_1}(q_2)$, and since $\frac{q_1}{q_2}u$ is centred around $q_2$ we have that 
$\frac{q_1}{q_2}u \in B_{q_2}$. Therefore $u\mapsto\frac{q_1}{q_2}u$ is a well--defined continuous bijection from $B_{q_1}$ to 
$B_{q_2}$ whose inverse map $v\mapsto \frac{q_2}{q_1}v$ is well--defined by the same arguments. 

\vspace{0.20in}

We denoted the map in the previous proposition by $\tau^{(-1)}$ since it satisfies the duality relation 
\begin{eqnarray*}
\langle \tau^{(1)} u,\tau^{(-1)}v \rangle_{q_2} &=& 
\left\langle u-\int_{\Omega}\!uq_2d\mu,
\frac{q_1}{q_2}v \right\rangle_{q_2} \\
&=& \int_{\Omega} u \frac{q_1}{q_2}v q_2d\mu - 
\left(\int_{\Omega}uq_2d\mu\right)\int_{\Omega}\frac{q_1}{q_2}v q_2d\mu \\
&=& \int_{\Omega} uv q_1d\mu \\
&=& \langle u,v \rangle_{q_1}, \quad \forall u,v \in B_{q_1},
\end{eqnarray*}
where the third equality follows from the fact that $v$ is centred around $q_1$.

\vspace{0.2in}

Let us now reflect on the collective results of Propositions \ref{mixtureg} to \ref{iso}. Proposition \ref{mixtureg} tells us that the convex mixture of two probability densities in the same ${\cal U}_p$ remains in the connected component ${\cal E}_p$ of ${\cal M}$, but not necessarily in the same neighbourhood.  Proposition \ref{extended_mix} then characterizes exactly those pairs $q_1$ and $q_2$ for which the convex mixture can be extended beyond the extreme points while remaining in the same connected component ${\cal E}_p$. For such pairs, Proposition \ref{iso} gives an isomorphism of tangent spaces $\tau^{(-1)}$ which satisfies the duality relation 
\eqref{dual_def} with respect to the Fisher information. However, we refrain from calling $\tau^{(-1)}$ a parallel transport, since it might fail to be well--defined when the points $q_1$ and $q_2$ do not satisfy the conditions of Proposition \ref{iso}. Nevertheless, we can still compute the derivative of $\tau^{(-1)}$ along curves that satisfy these conditions, as is done in the next proposition.

\begin{proposition} Let $v\in T_p{\cal M}$ be a tangent vector at $p\in{\cal M}$ and $s \in S(T{\cal M})$ be a differentiable vector field. If 
there exist a differentiable curve $\gamma :(-\varepsilon ,\varepsilon)\rightarrow
{\cal M}$ such that $p=\gamma (0)$, $v=\gamma'(0)$, and whose image consists entirely of points 
satisfying the hypotheses of Proposition \ref{iso}, then
\begin{equation}
(D^{(-1)}_vs)(p):=(d_v s)(p) + s(p)\ell^\prime(0)\in B_p,
\label{mix}
\end{equation}
where $(d_v s)(p)$ denotes the directional derivative of $s$ in the direction of $v$ in the Banach space $L^{\Phi_1}(p)$, and 
$\ell(t)=\log(\gamma(t))$. 
\label{mix_prop}
\end{proposition}
{\em Proof:} For $h$ sufficiently small, it follows from Proposition \ref{iso} that 
$\tau^{(-1)}_{\gamma(h)\gamma(0)}s(\gamma(h))\in B_{\gamma(0)}$. The result then follows from the following calculation:
\begin{eqnarray*}
\lim_{h\rightarrow 0} \frac{1}{h}\left[\tau^{(-1)}_{\gamma(h)\gamma(0)}
s(\gamma(h))-s(\gamma(0))\right] 
&=& \lim_{h\rightarrow 0} \frac{1}{h}\left[\frac{\gamma(h)}{\gamma(0)}
s(\gamma(h))-s(\gamma(0))\right] \\
&=& \lim_{h\rightarrow 0} \frac{1}{h}\left[s(\gamma(h))-s(\gamma(0))\right]+
\lim_{h\rightarrow 0} \frac{1}{h}\left[\frac{\gamma(h)-\gamma(0)}{\gamma(0)}
s(\gamma(h))\right] \\ 
&=& (d_vs)(p) + s(p)\ell^\prime(0).
\end{eqnarray*}

Despite satisfying all the usual properties of a covariant derivative, such as linearity and Leibniz rule, the differential operator 
$D_v^{(-1)}$ might fail to be well--defined when no curve satisfying the conditions of Proposition \ref{mix_prop} exists. For this reason, we simply call it the $(-1)$--derivative in the direction of those $v$ for which it is well--defined. In the next section, we will see that 
$D_v^{(-1)}$ is part of a one--parameter family of derivatives defined for exactly this class of tangent vectors.

\section{$\alpha$--derivatives}

\addtocounter{equation}{39}

In this section, we define an infinite--dimensional analogue of the 
$\alpha$-connections introduced in the parametric case independently in 
\citet{Chentsov82} and \citet{Amari85}. We use the same 
technique proposed in \citet{GibiliscoPistone98}, namely
exploring the geometry of spheres in the Lebesgue spaces $L^r$, but modified 
in such a way that the resulting derivatives all act on sections of the tangent bundle
$T{\cal M}$. The price we pay is that our derivatives are not defined for all tangent vectors, but only those satisfying the conditions of  Proposition 
\ref{mix_prop}.

We begin with the Amari-Nagaoka $\alpha$-embeddings
\begin{eqnarray}
\ell_{\alpha}:{\cal M} &\rightarrow&  L^r(\mu) \nonumber \\
p &\mapsto& \frac{2}{1-\alpha}p^{\frac{1-\alpha}{2}}, \quad \alpha \in [-1,1),
\end{eqnarray}
where $r=\frac{2}{1-\alpha}$.

Observe that 
\[\|\ell_{\alpha}(p)\|_r=\left[\int_{\Omega}\ell_{\alpha}(p)^rd\mu\right]^{1/r}
=\left[\int_{\Omega}\left(\frac{2}{1-\alpha}p^{\frac{1-\alpha}{2}}\right)^rd\mu
\right]^{1/r}=r,\]
so $\ell_{\alpha}(p) \in S^r(\mu)$, the sphere of radius $r$ in $L^r(\mu)$
(we warn the reader that, throughout this paper, the $r$ in $S^r$ refers to the
fact that this is a sphere of radius $r$, while the fact that it is a subset 
of $L^r$ is judiciously omitted from the notation).

According to \citet{GibiliscoPistone98}, the tangent space
to $S^r(\mu)$ at a point $f$ is 
\begin{equation}
T_fS^r(\mu)=\left\{g \in L^r(\mu):\int_{\Omega}gf^*d\mu=0\right\},
\end{equation}
where $f^*=\mbox{sgn}(f)|f|^{r-1}$. In our case,
\begin{equation}
f=\ell_{\alpha}(p)=rp^{1/r}
\end{equation}
so that
\begin{equation}
f^*=\left(rp^{1/r}\right)^{r-1}=r^{r-1}p^{1-1/r}.
\end{equation}
Therefore, the tangent space to $S^r(\mu)$ at $rp^{1/r}$ is
\begin{equation}
T_{rp^{1/r}}S^r(\mu)=\left\{g \in L^r(\mu):\int_{\Omega}gp^{1-1/r}d\mu= 0\right\}.
\end{equation}

We now look for a concrete realization of the push-forward of the map 
$\ell_{\alpha}$ when the tangent space $T_p{\cal M}$ is  identified with
$B_p$ as in the previous sections. Since
\[\frac{d}{dt}\left(\frac{2}{1-\alpha}p^{\frac{1-\alpha}{2}}\right)=
p^{\frac{1-\alpha}{2}}\frac{d\log p}{dt},\]
the $\alpha$--push-forward can be formally implemented as
\begin{eqnarray}
(\ell_{\alpha})_{*(p)}:{T_p{\cal M}=B_p} &\rightarrow& T_{rp^{1/r}}S^r(\mu) \nonumber \\
u &\mapsto& p^{\frac{1-\alpha}{2}}u.
\end{eqnarray}
For this to be well defined, we need to check that $p^{\frac{1-\alpha}{2}}u$
is an element of $T_{rp^{1/r}}S^r(\mu)$. Indeed, since $L^{\Phi_1}(p)\subset
L^s(p)$ for all $s\geq1$, we have that
\[\int_{\Omega}\left|p^{1/r}u\right|^rd\mu=\int_{\Omega}|u|Ä^rpd\mu < \infty,\]
so $p^{\frac{1-\alpha}{2}}u \in L^r(\mu)$. Moreover
\[\int_{\Omega}p^{1/r}up^{1-1/r}d\mu=\int_{\Omega}upd\mu=0,\]
which verifies that $p^{1/r}u \in T_{rp^{1/r}}S^r(\mu)$.

The sphere $S^r(\mu)$ inherits a natural connection obtained by projecting
the trivial connection on $L^r(\mu)$ (the one where parallel transport is 
just the identity map) onto its tangent space at each point. For each
$f \in S^r(\mu)$, a canonical projection from the tangent space 
$T_fL^r(\mu)$ onto
the tangent space $T_fS^r(\mu)$ can be uniquely defined, since the spaces 
$L^r(\mu)$ are uniformly convex (see \citet{GibiliscoIsola99}), and is given by
\begin{eqnarray}
\Pi_f:T_fL^r(\mu) &\rightarrow& T_fS^r(\mu) \nonumber \\
g &\mapsto& g-\left(r^{-r}\int_{\Omega}gf^*d\mu\right)f.
\end{eqnarray}
When $f=rp^{1/r}$ and $f^*=r^{r-1}p^{1-1/r}$, the formula above gives
\begin{eqnarray}
\Pi_{rp^{1/r}}:T_{rp^{1/r}}L^r(\mu) &\rightarrow& T_{rp^{1/r}}S^r(\mu) \nonumber \\
g &\mapsto& g-\left(\int_{\Omega}gp^{1-1/r}d\mu\right)p^{1/r}.
\end{eqnarray}

We are now ready to introduce the $\alpha$--derivative. Suppose that $\gamma :(-\varepsilon ,\varepsilon)\rightarrow{\cal M}$ is a smooth curve whose image consists entirely of points satisfying the conditions of Proposition \ref{iso}. Then the $\alpha$--push--forward of an arbitrary vector field 
$s \in S(T{\cal M})$ along $\gamma$ is  
\[(\ell_{\alpha})_{*(\gamma(t))}s=\gamma(t)^{1/r}s(\gamma(t)),\]
while the $\alpha$--push--forward of the tangent vector $v=\dot\gamma(0)\in T{\cal M}_p$ is 
\[(\ell_{\alpha})_{*(p)}v=p^{1/r}v.\]
Therefore, the covariant derivative of $(\ell_{\alpha})_{*(\gamma(t))}s(\gamma(t))$ in the direction of 
$(\ell_{\alpha})_{*(p)}v$ with respect to the trivial connection $\widetilde\nabla$ on $L^r(\mu)$ is given by
\begin{eqnarray*}
\widetilde{\nabla}_{(\ell_{\alpha})_{*(p)}v}(\ell_{\alpha})_{*(\gamma(t))}s&=&
\frac{d}{dt}\left.\left(\gamma(t)^{1/r}s(\gamma(t))\right)\right|_{t=0}\\
&=&\frac{1}{r}p^{1/r}\left.\frac{d\log(\gamma(t))}{dt}\right|_{t=0}s(p)+p^{1/r}\left.\frac{ds(\gamma(t))}{dt}\right|_{t=0} \\
&=& \frac{1}{r}p^{1/r}\ell'(0)s(p)+p^{1/r}\left(d_vs\right)(p).
\end{eqnarray*}
Using the projection $\Pi_{rp^{1/r}}$ to obtain a tangent vector in 
$T_{rp^{1/r}}S^r(\mu)$ we get
\[
\Pi_{rp^{1/r}}\widetilde{\nabla}_{(\ell_{\alpha})_{*(p)}v}(\ell_{\alpha})_{*(\gamma(t))}s=
p^{1/r}\left[\frac{1}{r}\ell'(0)s(p)+(d_vs)(p)-
\int_{\Omega}\left(\frac{1}{r}\ell'(0)s(p)+
(d_vs)(p)\right)pd\mu\right].\]
It then follows from Proposition \ref{mix_prop} that $\frac{1}{r}\ell'(0)s(p)+(d_vs)(p)\in L^{\Phi_1}(p)$, which implies that the expression above belongs to the image of $(\ell_{\alpha})_{*(p)}$. Therefore, we can pull it back to $T_p{\cal M}$ using 
$(\ell_{\alpha})_{*(p)}^{-1}$, from which we obtain 
\begin{eqnarray}
(\ell_{\alpha})_{*(p)}^{-1}
\left[\Pi_{rp^{1/r}}\widetilde{\nabla}_{(\ell_{\alpha})_{*(p)}v}
(\ell_{\alpha})_{*(\gamma(t))}s\right]&=& \frac{1}{r}\ell'(0)s(p)+(d_vs)(p) \nonumber \\
&&-
\int_{\Omega}\left(\frac{1}{r}\ell'(0)s(p)+
(d_vs)(p)\right)pd\mu.
\label{alpha_def_aux}
\end{eqnarray}
As this construction shows, the $\alpha$--derivatives can be rigorously defined on the tangent bundle $T{\cal M}$ as follows:

\begin{definition} For $\alpha \in [-1,1)$, let $\gamma :
(-\varepsilon ,\varepsilon)\rightarrow
{\cal M}$ be a smooth curve such that $p=\gamma (0)$ and whose image consists  
entirely of points satisfying the conditions of Proposition \ref{iso}. The 
$\alpha$--derivative of a differentiable vector field $s\in S(T{\cal M})$ in the direction of $v=\gamma^\prime(0)$ is given by
\begin{equation}
\left(D^{\alpha}_vs\right)(p)=(\ell_{\alpha})_{*(p)}^{-1}
\left[\Pi_{rp^{1/r}}\widetilde{\nabla}_{(\ell_{\alpha})_{*(p)}v}
(\ell_{\alpha})_{*(\gamma(t))}s\right].
\label{alpha}
\end{equation}
\label{def_alpha}
\end{definition}

Before we proceed, observe that since $s(\gamma(t)) \in B_{\gamma(t)}$ for each
$t \in (-\varepsilon,\varepsilon)$, we have
\begin{eqnarray*}
\frac{d}{dt}\int_{\Omega}s(\gamma(t))\gamma(t)d\mu &=& 0 \\
\int_{\Omega}\frac{ds(\gamma(t))}{dt}\gamma(t)d\mu &=& 
-\int_{\Omega}s(\gamma(t))\frac{d\gamma(t)}{dt}d\mu \\
\int_{\Omega}\frac{ds(\gamma(t))}{dt}\gamma(t)d\mu &=& 
-\int_{\Omega}s(\gamma(t))\frac{d\log(\gamma(t))}{dt}\gamma(t)d\mu.
\end{eqnarray*}
In particular, for $t=0$, we get
\begin{equation}
\int_{\Omega}\left(d_vs\right)(p)pd\mu = 
-\int_{\Omega}s(p)\dot{\ell}(0)pd\mu
\label{log}
\end{equation}
Inserting this relation into \eqref{alpha_def_aux} with $\alpha=-1$, corresponding to $r=1$, leads to
\begin{equation}
\left(D^{(-1)}_vs\right)(p)= (d_v s)(p) + 
s(p)\ell'(0),
\end{equation}
which coincides with \eqref{mix}. 

Recall that the covariant derivative associated with the exponential parallel transport \eqref{exp_parallel} was computed in  
 \citet[proposition 25]{GibiliscoPistone98} as  
\begin{equation}
\left(\nabla_v^{(1)}s\right)(p)=(d_v s)(p) - \int_{\Omega}(d_v s)(p)pd\mu.
\label{exp}
\end{equation}
The next proposition shows that the relation between the exponential connection and the $\alpha$--derivatives just defined is the same as in the 
parametric case. Its proof resembles the calculation in the last pages of 
\citet{GibiliscoPistone98}, except that all our derivatives act on the same 
bundle, whereas in \citet{GibiliscoPistone98} each one is defined on its
own bundle-connection pair. 

\begin{proposition} The exponential connection and the $\alpha$--derivatives on 
$T{\cal M}$ satisfy
\begin{equation}
D^{\alpha}=\frac{1+\alpha}{2}\nabla^{(1)}+\frac{1-\alpha}{2}D^{(-1)}.
\end{equation}
\end{proposition}
{\em Proof:} Let $\ell(t)=\log (\gamma(t))$ with $\gamma$, $s$, $p$ and $v$
as in definition \ref{def_alpha}. Inserting \eqref{log} into \eqref{alpha} gives 
\begin{eqnarray*}
\left(D^{\alpha}_vs\right)(p) &=& \frac{1}{r}\ell'(0)s(p)+(d_vs)(p) + 
\left(\frac{1}{r} - 1\right)\int_{\Omega}(d_vs)(p)pd\mu  \\
&=& \left(\frac{1+\alpha}{2}\right)\left[(d_v s)(p) - 
\int_{\Omega}(d_v s)(p)\right] + \left(\frac{1-\alpha}{2}\right)
[(d_v s)(p) + 
s(p)\ell'(0)] \\
&=& \frac{1+\alpha}{2}\left(\nabla_v^{(1)} s\right)(p) +
\frac{1-\alpha}{2}\left(D_v^{(-1)} s\right)(p).
\end{eqnarray*}

\section{Auto-parallel Curves} 

\addtocounter{equation}{53}

We now investigate some of the auto--parallel curves associated with the derivatives introduced in the previous sections. First observe that  a one--dimensional exponential model
of the form
\[q(t)=\frac{e^{tu}}{Z_p(tu)}p,\qquad u\in B_p, \quad t\in (-\varepsilon, \varepsilon),\]
which obviously belong to the connected component ${\cal E}_p$, is an auto--parallel curve for $\nabla^{(1)}$, since its tangent vector field 
$s(t)=\frac{d}{dt}\left(\log\frac{q(t)}{p}\right)$ (according to \citet[proposition 21]{PistoneRogantin99}) satisfies 
\begin{eqnarray*}
\left(\nabla_{\dot{q}(t)}^{(1)}s(t)\right)(q(t)) &=&
\frac{d^2}{dt^2}\left(tu - \log Z_p(tu)\right) - E_{q(t)}\left(
\frac{d^2}{dt^2}\left(tu - \log Z_p(tu)\right)\right) \\
&=& -\frac{d^2}{dt^2}\log Z_p(tu) + E_{q(t)}\left(\frac{d^2}{dt^2}
\log Z_p(tu)\right)= 0.
\end{eqnarray*}

Next observe that for $q_1$ and $q_2$ satisfying the conditions of Proposition \ref{iso}, a mixture model of the form   
\[q(t)=tq_1 + (1-t)q_2, \qquad q_1,q_2\in {\cal U}_p, \quad t \in [0,1],\]
which belongs to the connected component ${\cal E}_p$ according to Proposition \ref{mixtureg}, is an auto--parallel 
curve for $D^{(-1)}$, since the tangent vector field 
$s(t)=\frac{d}{dt}\left(\log\frac{q(t)}{p}\right)$ satisfies 
\begin{eqnarray*}
\left(D_{(e^{-1}_p\circ q)'(t)}^{(-1)}s(t)\right)(q(t)) &=&
\frac{d^2}{dt^2}\left[\log\frac{tq_1+(1-t)q_2}{p}\right] \\
& & \qquad\qquad  + 
\frac{d}{dt}\left[\log\frac{tq_1+(1-t)q_2}{p}\right]\frac{d}{dt}
[\log tq_1+(1-t)q_2] \\
&=& \frac{d}{dt}\left[\frac{p}{tq_1+(1-t)q_2}\frac{(q_1-q_2)}{p}
\right]\\
& & \qquad\qquad
+\left(\frac{p}{tq_1+(1-t)q_2}\frac{q_1-q_2}{p}\right)
\frac{(q_1-q_2)}{tq_1+(1-t)q_2} \\
&=& -\left(\frac{(q_1-q_2)}{tq_1+(1-t)q_2}\right)^2 +
\left(\frac{(q_1-q_2)}{tq_1+(1-t)q_2}\right)^2 =0.
\end{eqnarray*}

The next theorem establishes the corresponding result for the $\alpha$--derivatives. 

\begin{proposition} For $\alpha\in(-1,1)$, the $\alpha$--auto--parallel curves between two of points $q_1$ and $q_2$ in ${\cal U}_p$ satisfying the conditions of Proposition \ref{iso}, for some $p\in {\cal M}$, belongs to the connected component ${\cal E}_p$. 
\end{proposition}
{\em Proof:} Using the same notation as in Proposition \ref{mixtureg},  we have that the $\alpha$--auto--parallel curve connecting $q_1,q_2 \in {\cal E}(p)$ is the pull back of the arc of great circle connecting their images $f_1=\ell_\alpha(q_1)$ and $f_2=\ell_\alpha(q_2)$ on the sphere $S^r(\mu)$. Now if $t f_1 + (1-t)f_2$ is the straight line connecting $f_1$ and $f_2$ in $L^r(\mu)$, then for each fixed $t\in[0,1]$ the corresponding point on the sphere of radius $r$ is 
\begin{equation}
f(t)=\frac{r}{k(t)}[ t f_1 + (1-t)f_2],
\end{equation}
where $k(t)=\|t f_1 + (1-t)f_2\|_r$. Let us write its inverse image with respect to the $\alpha$--embedding as
\[e^{\tilde u}p=(\ell_\alpha)^{-1}(f(t))=\frac{1}{k(t)^r}[tf_1 +(1-t)f_2]^r,\]
for some random variable $\tilde u$. Following the argument in proposition \ref{mixtureg}, we see that
\begin{eqnarray*}
e^{\beta\tilde u}&=&\frac{1}{p^\beta k(t)^{\beta r}}[tf_1 +(1-t)f_2]^{\beta r} \\
&\leq& \left[\frac{2r}{k(t)}\right]^{\beta r}[t^{\beta r}e^{\beta \tilde u_1}+(1-t)^{\beta r}e^{\beta \tilde u_2}],
\end{eqnarray*}
so that 
\begin{equation}
\label{est1}
\int_\Omega e^{\beta \tilde u}pd\mu < \infty,
\end{equation}
since both $\beta u_1$ and $\beta u_2$ are in $\tilde L^{\Phi_1}(p)$. Furthermore, 
\[e^{-\beta\tilde u}= \frac{p^{\beta} k(t)^{\beta r}}{[tf_1 +(1-t)f_2]^{\beta r}} \leq \left[\frac{k(t)}{t r}\right]^{\beta r}
e^{-\beta \tilde u_1},\]
so that
\begin{equation}
\label{est2}
\int_\Omega e^{-\beta \tilde u}pd\mu < \infty,
\end{equation}
which together with \eqref{est1}, imply that $\tilde u\in {\cal Z}_p$. To complete the proof we can define
\begin{equation}
u=\tilde u - \int_\Omega \tilde u pd\mu,
\end{equation}
to obtain that $u\in B_p$ and
\begin{equation}
(\ell_\alpha)^{-1}(f(t))=\frac{e^u}{Z_p(u)}p \in {\cal E}(p).
\end{equation}

\section{Further developments}

We have seen that using $L^{\Phi_1}$ as the coordinate space for the infinite--dimensional
information manifold leads to a well--defined isomorphism $\tau^{(-1)}$ between the tangent spaces $B_{q_1}$ and $B_{q_2}$ whenever 
the difference of their log-likelihoods $u_1$ and $u_2$ is bounded. Moreover, this isomorphism is dual to the exponential parallel transport with respect to the generalized Fisher metric. The next step in our program is to show that the Kullback-Leibler relative entropy is the statistical divergence associated
with the dualistic triple $(\tau^{(1)},\tau^{(-1)},g)$ (see \citet{AmariNagaoka00}). In the same vein,
since our interpolation family of $\alpha$--derivatives satisfy the same convex mixture structure as in 
finite dimensions, we are led to the study of the infinite--dimensional analogues of the $\alpha$--divergences. The completion of this
circle of ideas would be an infinite--dimensional generalization of the projection theorems obtained by Amari
in the finite dimensional case. Namely, one seeks to prove that, given a point $p\in{\cal M}$ and an $\alpha$--flat
submanifold ${\cal S}$, then the point $q\in{\cal S}$ with minimal $\alpha$-divergence from $p$ is obtained by
projecting $p$ orthogonally (with respect to the Fisher metric) onto ${\cal S}$ following a $-\alpha$--geodesic. An
equally ambitious result to be pursued is the infinite--dimensional analogue of Centsov's theorem, which would
characterise the generalized Fisher metric as the unique continuous scalar product on ${\cal M}$ which is reduced
by Markov morphisms on the tangent space. 

\acknowledgements
I would like to acknowledge the financial suport from CAPES, Brazil, and of an ORS 
scholarship from the British government received while I was a PhD student at King's College London, where
this research was initiated. It is a pleasure to thank R.F. Streater for suggesting the theme and 
thoroughly discussing the results obtained. I am also grateful to
P. Gibilisco and G. Pistone for estimulating discussions during the Information Geometry and Applications 
conference 
held in Pescara, July 2002.  Finally, I am indebted to H. Nagaoka for pointing out a crucial mistake in the previous version of this paper, and to two anonymous referees for marked improvements leading to the current version. 

\def\cprime{$'$}

\end{document}